\documentstyle[epsf,12pt]{article}
\begin{document}
\begin{center}
\Large{\bf Possible Striking Signals for a Quark-Gluon Plasma at RHIC}\\
\large{S.J. Lindenbaum$^{a,b}$ and R.S. Longacre$^a$\\
$^a$Brookhaven National Laboratory, Upton, NY 11973, USA\\
$^b$City College of New York, NY  10031, USA\footnote{This research was 
supported by the U.S. Department of Energy under Contract No. 
DE-AC02-98CH10886 and the City College of New York Physics Department}}
\end{center}

\begin{abstract}
We believe that one can have serious reservations as to whether heavy 
ion collisions (e.g. 100 GeV/n Au + 100 GeV/n Au) can lead to Thermal 
and Chemical equilibrium over large regions (particularly if it is 
assumed this happens whenever QGP is produced at RHIC-that is if it 
is produced).

It is at present not clear that the collision dynamics and times
available will lead to this. An alternate scenario proposed by Van Hove 
where localized in rapidity bubbles of plasma may well be more probable,  
and may well occur at least some of the time, and some of the time mainly 
survive to the final state. If this occurs we have developed a series of 
event generators to extend and describe these phenomena. A Van Hove
type[6,7] spherical bubble at $\eta=0$ is embedded in a resonable event
generator in qualitative agreement with Hijing etc[12].  The plasma bubble
hadronized at a temperature of 170 Mev according to the model developed
by Koch, M\"{u}ller and Rafelski[21].  The amount of available energy assumed 
in the bubble is selected by that in a small central circular cross-section
of radius $\approx 1.3fm$ or $\approx 2.5fm$ in 100 Gev/n Au+AU, central 
events. The results predict (with the assumptions stated, possible) Striking 
Signals which may allow strong evidence for a QGP which cannot be explained by 
alternative conventional physics arguments, and thus may be crucial elements 
in establishing a QGP. We are also applying these techniques to investigating  
Kharzeev and Pisarski bubbles of metastable vacua with odd CP.
\end{abstract}

\section{Introduction}

For about two decades there has been considerable interest in the
possibility that at sufficiently high temperatures or baryon density one
or more phase transitions will lead to formation of a quark-gluon plasma
characterized by deconfinement, and perhaps chiral symmetry restoration
[1,2] Very high energy heavy ion collisions was the community consensus as the
best hope of forming a quark- gluon plasma {QGP} and this led to RHIC
[3] with its up to 100 GeV/nucleon Au colliding with 100 GeV/nucleon Au
colliding beam accelerator.

RHIC is expected to possibly begin to provide experimental data in
2000. Therefore it is timely to ask the question what could some
striking signals possibly be generated and detected which would provide very
strong evidence, hopefully convincing, for a QGP.

Of course it may be that QCD is not correct in this new energy domain,
or that lattice gauge calculations and phenomenological models used to
predict the formation and characteristics of a QGP are not applicable to the
real dynamical situation.

\section {Formation of a Quark Gluon Plasma}

The first question one must address is that even if present ideas
on the conditions necessary for formation of a QGP are approximately
correct, does the actual dynamical situation at RHIC allow them to be 
achieved and if so how.

We have from the early days of RHIC planning, had serious reservations
in regard to this[4,5]. Although many theoretical calculations assume that
central collisions of heavy ions can be described by employing local
thermodynamic and chemical equilibrium which adjusts adiabatically as the 
collision zone develops in space and time, one can have serious reservations 
as to whether the collision dynamics allow and whether sufficient time 
exists for reaching such overall thermodynamic and chemical equilibrium.

A second approach has been to recognize that it is unlikely that
thermalized, conditions can describe the whole collision dynamics in 
particular the phase  transition itself, and thus if new phenomena 
(QGP) etc. occur, they form under inherently non equilibrium conditions. 
This scenario has been suggested and strongly emphasized by Van Hove[6,7].

Van Hove's scenario would lead to local droplets of QGP if QGP is
formed. As the droplets expand they could, in some cases separate into 
smaller droplets. It was expected that these QGP droplets could hadronize 
by deflagration since this appeared to be the more likely of the two
possible explosive phenomena as it was favored by entropy considerations
[6-8].

It should be noted that these non-equilibrium treatments have assumed
that the chemical potential is zero (i.e. baryon no. $B\approx 0$) and 
thus are directly applicable to the central region. At RHIC 100 Gev/n Au on 
100 Gev/n Au central events are expected to have B close enough to zero for 
our calculational purposes, especially in the Quark-gluon plasma bubbles.
It is expected that the central region at RHIC will have $B\approx 0$, 
however Van Hove's approach may even be qualitatively correct for the 
behavior of plasma droplets originally formed even in baryon dense regions, 
since they rely mainly on the existence of a large amount of latent heat 
and latent entropy in the phase transition, conditions which also apply 
to baryon dense regions.

If plasma droplets (possibly after breaking up into two or more droplets)
hadronize by deflagration, Van Hove's scenario concluded that the
resulting distribution of hadrons should show maxima approximately at the
rapidities of the droplets. The expected width of the maxima was 
estimated to be $\sim 1$ rapidity unit and have angular distributions 
characteristic of a deflagration occurring in the plasma droplets.  He 
also expected the generally expected plasma signals such as enhanced 
strange and multistrange particles, enhanced anti particles, enhanced 
lepton pair production, etc.

If QGP is formed it may be easier to conceive of Van Hove's local
droplet scenario occurring since the long times required for overall
thermodynamic and chemical equilibrium to be attained may not be attainable.

If bubbles in the VAN HOVE sense are not created, the techniques
described in this paper will still be applicable if localized in rapidity 
regions, chunks of QGP are created by any mechanism and the hadronic 
signals from them are not too degraded by subsequent interactions.  
Of course if many droplets (or chunks of rapidity localized QGP) are 
formed over large regions one might approximate this situation by 
assuming thermodynamic and chemical equilibrium over a large region. 
It should also be noted, as we will be discussing, that the droplets 
(or rapidity localized) chunks of QGP will be expected to generate 
very striking signals which are detectable and could provide 
convincing evidence for a QGP.

In this regard one should note that a bubble or rapidity localized
region in QGP will be surrounded by non-QGP background, and thus allow 
a comparison of QGP and non-QGP backgrounds in the same event.  This 
is a potent way to possibly establish a QGP. One should also note that 
these bubble like events only need to happen and their characteristics 
survive some of the time (even rarely) provided they provide sufficient
statistics in order to provide powerful evidence for a QGP.

\section {Event Generation}

One must use an event generator suitable for estimating the non plasma
events, and as a second step embed the plasma bubbles formation and
deflagration, and compare plasma regions in the event with the non-plasma 
regions, and then also compare non plasma events with the plasma 
events to determine the detectability of plasma bubbles, and whether 
they would be striking enough to at least provide credible evidence 
for plasma formation, or hopefully even convincing evidence for a QGP.

\section {RHIC::EVENT - a non plasma generator}

The geometry of the A-A collision is taken into account by populating
nucleons in the target and the projectile systems according to a
Woods-Saxon Distribution. The nucleons have a Gaussian distribution of  
Fermi momentum with a small $\sigma$ of 200 MeV/c. The nucleons are off 
the mass shell, with an average binding energy of 8 MeV. A collision 
interaction between a projectile, and target nucleon occurs when the 
distance of closest approach is less than $\sqrt \sigma/\pi$ with 
$\sigma$=33 mb. Thus this is 1.025 Fermi-the approximate size of the 
proton.  The center of mass system of the $N-N$ collision is used.

ISAJET[9] - an event generator for high energy $N-N$ interactions, has
been successful in explaining these interactions, and is therefore
used as the basis of building RHIC:: EVENT. A high energy 
Nucleus+Nucleus (i.e. A+A) collision is represented as resulting from 
a series of nucleon-nucleon collisions but taking into account  that 
after a nucleon suffers a collision it becomes a forward going diquark.

The MINBIAS routine of ISAJET, is based on inclusive high energy $N-N$
interactions forming multi-pomeron chains, with each chain fragmenting
according to the Field-Feynman algorithm[10].

MINBIAS is used to compute the energy loss of the colliding nucleons and
the produced particles . Unlike earlier event generators based on ISAJET
(i.e. HIJET[11]) instead of using the leading baryon for the diquark for 
the next collision, RHIC::EVENT sums up the momentum of the particles 
produced by MINBIAS that one wants to associate with the diquark cluster 
and RHIC::EVENT does not allow the leading diquark to change flavors. It 
is  speculated that the forward going cluster is what the fragmentation 
region  is. The larger the cluster size and momentum the more RHIC::EVENT 
creates  nuclear transparency. This algorithm can generate a very nice
flat central region plateau for RHIC energies which is flat over $\approx$
5 units of rapidity.

A popular event generator, HIJING[12], is a Monte Carlo model which 
combines Fritiof[13] for soft beam jet fragmentation, and Pythias[14] for 
semi-hard mini jet Physics[15].  Gyulassy[15] shows in his Fig.~1 that 
HIJING reasonably represents  the number of charged particles per unit 
$y$ or $\eta$ as a function of $y$ or $\eta$ for central-i.e. very high 
multiplicity events. When HIJING is applied to 100 GeV/nucleon Au 
colliding with 100 GeV/nucleon Au Gyulassy shows that for ``central events'' 
at RHIC (his Fig.~1) a flat plateau is formed centered around midrapidity 
or $\eta$ and extends (within $\sim$20\% from the peak value) for a total 
of approximately four $y$ or $\eta$ units, and thus is qualitatively in 
agreement with the results of our RHIC::EVENT generator with leading 
cluster equal to only the diquark. Gyulassy points out that HIJING 
results are similar to those obtained with other models[13-20].

     It should be noted that uncertainty originates from the poorly
known early evolution of the mini-jet plasma, and this affects the
height of the plateau in the calculations of Gyulassy[15].  Computing
the early evolution of the color fields more reliably will allow better
estimation of the plateau height.

\section{RHIC::PLASMA - a spherical plasma bubble(s)\\ generator}

Plasma formation and decay is a very uncertain process and our first
model of this phenomenon is based on work of L. Van Hove[6,7]. Van Hove's 
model predicted that the deconfinement transition is described in terms 
of formation of QCD strings in the expanding plasma. These strings in 
the plasma are stopped from expanding because of string tension and 
because of string breaking droplets or bubbles are formed. These 
bubbles hadronize by deflagration at the phase transition through their 
outer surface by ejection of low pressure hadron gas with velocities 
associated with the critical temperature. This process creates a 
rapidity distribution $dn/dy$ or a pseudorapidity distribution 
$dn/d\eta$ of hadrons with isolated maxima of width $\delta y$ or 
$\delta\eta\sim 1$ or in our case high bumpy regions $\sim 2$ units 
of rapidity wide on an event-by-event basis.

The generator RHIC::PLASMA takes the regular RHIC::EVENT generation and
Van Hove type[6,7] spherical plasma bubbles generated by a regional 
tagging scheme which takes particles from the intial RHIC::EVENT 
collisions and converts their energy and baryon number to a plasma 
bubble. The tagging region is defined as a circular cross sectional 
area located within the center of the overlap region with radius $R_{tag}$.
Each region is then subdivided in phase space by a longitudinal momentum 
cuts (three bubbles can be defined in present code). The bubble has an 
intial energy and baryon number, where its rest frame is given by the 
total momentum energy four vector of the sum of the tagged particles 
that make up the bubble. The thermal dynamics of the bubble is generated 
in accord with the model of Koch, Muller and Rafelski [21]. Initially the 
tagged volume is used to define the chemical potential and thus the quark 
and gluon populations for a plasma sitting at its critical temperature. 
Reference [21] has worked out these equations up to a temperature of 170 
MeV. We expand the volume until we get a self consistent condition that 
satisfies energy and baryon number between the tagged particles and the 
plasma bubble. This self consistent condition is equal to the number of 
quarks, anti-quarks and gluons ($N_qN_{\bar q}N_g$). $N_s$ and $N_{\bar s}$ 
are the strange quark and anti-quark numbers where $N_q$ and $N_{\bar q}$ 
are the light quarks ($u,d$). When hadronization takes place gluons 
fragment into quarks and anti-quarks leading to an effective number given 
by $\tilde{N}_q=N_q+f_qN_g$; $\tilde{N}_{\bar q}=N_{\bar q}+f_qN_g$;
$\tilde{N}_s=N_s+f_sN_g$; $\tilde{N}_{\bar s}=N_{\bar s}+f_sN_g$, where
$f_q=$ gluon fragmentation function for $u$ and $d$ quarks and $f_s=$
gluon fragmentation function for $s$ quarks. The average number of
particles produced each with a Boltzman distribution corresponding to
the critical temperature is:
$N_\pi = \alpha\tilde{N}_q\tilde{N}_{\bar q}$,
$N_K =\alpha\tilde{N}_q\tilde{N}_{\bar s}$,
$N_\phi =\alpha\tilde{N}_s\tilde{N}_{\bar s}$,
$N_n={1\over{3!}}\beta\tilde{N}_q\tilde{N}_q\tilde{N}_q$,
$N_{\bar n}={1\over{3!}}\beta\tilde{N}_{\bar q}\tilde{N}_{\bar q}
\tilde{N}_{\bar q}$,
$N_\Lambda ={1\over{2!}}\beta\tilde{N}_q\tilde{N}_q\tilde{N}_s$,
$N_{\bar\Lambda} ={1\over{2!}}\beta\tilde{N}_{\bar q}\tilde{N}_{\bar q}
\tilde{N}_{\bar s}$,
$N_\Xi ={1\over{2!}}\beta\tilde{N}_q\tilde{N}_s\tilde{N}_s$,
$N_{\bar\Xi} ={1\over{2!}}\beta\tilde{N}_{\bar q}\tilde{N}_{\bar s}
\tilde{N}_{\bar s}$,
$N_\Omega ={1\over{3!}}\beta\tilde{N}_s\tilde{N}_s\tilde{N}_s$,
$N_{\bar\Omega} ={1\over{3!}}\beta\tilde{N}_{\bar s}\tilde{N}_{\bar s}
\tilde{N}_{\bar s}$.

\noindent{$\alpha$} and $\beta$ are given by the equations:
\begin{eqnarray}
\alpha  & = & {4Q_1\over{\left( 3Q^2_1 + Q^2_2\right )}};\\
\beta  & = & {8\over{\left( 3Q^2_1 + Q^2_2\right )}};
\end{eqnarray}

\noindent{where} $Q_1=\tilde{N}_q+\tilde{N}_{\bar q}+\tilde{N}_s+
\tilde{N}_{\bar s}$ and $Q_2=\tilde{N}_q-\tilde{N}_{\bar q}$. These
equations make sure baryon number is conserved and particles are formed
randomly out of the particle densities. The above particles then are
generated with a Boltzman distribution of a 170 Mev temperature
spherically in phase space. The spherical distribution creates a bump 
in rapidity or pseudorapidity of the type that Van Hove[6,7] predicted. 
However, the width of the bumps is about 2 units.  The geometry of the 
expansion of the plasma bubble or fireball is dependent on QCD and the 
collision dynamics.  Other cases will be considered subsequently.

\section {RHIC::LANDAU}

Although we have so far considered spherical bubbles, any reasonable
bubble shape can be incorporated in our plasma generator.  There is a
one-to-one correspondence between the bubble shape assumed and the
effect on the rapidity distribution and structure in it caused by the
bubble.  Given a particular observed rapidity distribution, assumed to
be due to plasma bubble(s) formation, we can also deduce general
characteristics of the shape of the plasma regions.  For example the
shapes of three other localized in rapidity QGP regions (or bubble 
shapes), are treated below.

The decision as to whether to attribute an experimentally observed
bubble-like phenomenon to plasma formation will depend on the behavior
and the correlation of the various possible plasma signals associated
with it (some of which are shown in the figures), other characteristics 
of the data, all at the time known facts, and very importantly the 
lack of a viable other conventional physics alternative.

In particle and nuclear physics the fireballs (not in any way shown to
contain QGP) that have  been observed,  have an additional one
dimensional or longitudinal expansion left over from their production 
mechanism (Landau fireball[22]). The program RHIC::LANDAU takes the 
regular RHIC::PLASMA program and replaces the spherical expansion of a 
QGP bubble by an expansion that is consistent with thermal fits done on 
SS collisions at 200 GeV/c per nucleon incident on a target[23]. The 
width of the rapidity peak distribution changes from a two unit spread 
to a four unit spread, where 50\% of the tagged energy goes into particle 
production and 50\% goes into longitudinal expansion. In the language of 
Van Hove the string stopping due to breaking is not complete, so that 
longitudinal expansion is left in the individual strings or bubbles.

\section {RHIC::SMOKE}

The plasma bubbles that we have generated give structure to the rapidity
or pseudo-rapidity spectrum of the produced particles. This structure is
directly related to the geometry of the expansion. The geometry which 
has been observed in experiments is the elliptical expansion of the 
Landau fireball, however, sharp structure in pseudo-rapidity has been 
observed in cosmic ray experiments[24]. We can make our plasma bubbles 
produce sharp structures by changing an elliptical or watermelon shape 
to a flat or pancake shape. We do this only as a phenomenological 
manipulation of our code. The end process is a ring of particles emitted 
at a fixed angle to the beam axis much like an expanding smoke ring. 
Since these particles have nearly a fixed angle they lead to sharp
pseudo-rapidity bumps of the intermittent type seen in cosmic ray data[24].

\section {RHIC::CHIRAL}

Another novel event structure seen in cosmic ray data[25]is the so-called
``centauro-like'' and ``anti-centauro-like'' fluctuations of charged and
neutral particles. A proposed explanation of these events relies on a 
chiral phase transition. Rajagopal and Wilczek[26] proposed a ``quench'' 
scenario in which the hadronic condensate after a phase transition is 
initially chirally symmetric, but its evolution is taken to follow 
classical equations of motion at zero temperature. The non-equilibrium 
dynamics of the chiral transition (using a linear $\sigma$ model to describe 
the collective chiral behavior) in relativistic heavy ion collisions 
yields  large disoriented chiral condensates. These disoriented chiral 
condensates lead to the non-Poisson distribution of the ``centauro-like''
and the ``anti-centauro-like'' events. The probability distribution of 
neutral pions $P(R3)$[27], where $R_3 = {N_{\pi^0}\over (N_{\pi^+} + 
N_{\pi^0} + N\pi^-)}$ , is equal to $P(R_3) = {1\over{\sqrt{R_3}}}$.  
In contrast, typical hadronic collisions produce a binomial distribution 
of $R3$ peaked at the isospin symmetric value of 1/3.

RHIC::CHIRAL generates the above like structures using the RHIC::PLASMA
bubble code but making the bubble hadronize into a 70 MeV temperature
pion gas with a $R_3$ distribution chosen to be that of the disoriented 
chiral condensate model. In a given Au-Au event each bubble has an 
independent $R_3$.

It is obvious that there is a relationship between structures in the
rapidity distribution and the geometry of the plasma bubbles, which
cause them. Thus if a certain type of structure is observed in the future
experiments, this will infer the geometry characteristics of the bubble
region (or whatever) caused it.

\section {Event Generation and Detection}

We believe the STAR Detector[28], for example, is suitable for
investigating our predictions.  Any statements about STAR are our own
estimates (also based on estimates of others) merely used for
comparison to our prediction purposes, and should, therefore, be taken
as an approximation.

The high multiplicity expected in STAR central (and some non central)
events coupled with the almost complete solid angle coverage over a 
substantial central rapidity range for 100 GeV per nucleon gold on 
gold down to p+p collisions opens up the possibility of observing 
single events well enough to decide whether equilibrium, non-equilibrium 
or strikingly different events occur individually. This eliminates the 
possibility of false conclusions being drawn in inclusive studies which 
may average over events in such a fashion as to eliminate their most 
striking and important characteristics. This could be a crucial element 
in establishing that a Quark-Gluon Plasma is produced at RHIC or 
observing other new phenomena which may even violate QCD.

The availability of as full solid angle particle identification as
practical over a substantial particle momentum range will be of 
considerable aid in this program.

We believe the study of single events on an event by event basis is a
most potent approach to search for a quark gluon plasma or whatever 
else nature reveals at RHIC ie perhaps new physics beyond QCD.

{\bf{Some results of our calculations for 100 GeV/nucleon AU colliding 
with 100 GeV/nucleon Au single event central collisions at RHIC:}}

The STAR detector is the most appropriate for our approach, hence the
following calculations were made for it using the geometric acceptance
of the central TPC detector, and particle identification, by $de/dx$
(ionization loss), and the planned Time-of-Flight (TOF) covering the 
central TPC. Efficiencies for tracking and particle identification 
were not included.

Figures 1-4 show the calculated results for a single RHIC::Event. 
The plasma induced rapidity bumps you see in the Figs.(5-8) are 
based on choosing $R_{tag}\approx 1.3 fm$ which leads to the very modest 
assumption that only 4.5\% of the available energy in a central collision  
is converted to QGP (as described in the event generator section 
RHIC::PLASMA).

Please note that if $R_{tag}$ were chosen to be $\approx 2.5~fm$, which
would lead to $\approx$15\% of the available energy being converted to 
QGP, the rapidity bump amplitudes over background would increase by about 
a factor of three and the statistics (i.e. number of particles in the 
rapidity  bump) would increase by about a factor of three (i.e. the bump 
height relative to the background and the number of particles in the 
bump would scale up approximately as the percentage of energy converted 
to QGP [that is increase by the factor (\% energy of QGP)/4.5\%.]  The 
figures for $R_{tag}$ $\approx 2.5~fm$ (figs. 9-13) which converts 
$\approx$15\% of the available energy to QGP show truly dramatic 
$dn/d\eta$ vs. $\eta$ peaks for single ``central'' events.

\section {RHIC::PLASMA - spherical bubbles}

(a) The Plasma bubble is produced at central rapidity by 100 GeV/n Au
colliding with 100 GeV/n Au (results figs. 5-8) in the region with
$R_{tag\\}\approx 1.3~fm$ which contains only approximately 4.5\% of the 
available energy. Even this conservative estimate will as, we can see 
from (Figs. 5-8), produce striking signals in these single events.

(b) The plasma bubble is produced at central rapidity where the
$R_{tag}\approx 2.5~fm$ which contains $\approx$15\% of the available
energy.  These assumptions lead to truly dramatic signals for single
events (see figs. 9-13).

(c) Figures (14-16 show the psuedorapicity distribution of dET/dNch (average
energy per charged particles) for the three cases considered.  The increase
in ET for the plasma bubbles is attributable to the fact that the plasma 
contains much larger percentages of heavy hadrons such as protons, anti 
protons, and charged kaons, that the cascade (non-plasma events), and the 
plasma hadronizes at 170 Mev. These results come right out of the model with
no extra assumptions.  We can generate an enormous number of detailed plots 
of desired quantities, but this is obviously not suitable for a publication, 
so that we must be selective.  However when data is obtained with rhic we 
will generate what is needed and relevant to confront our model, and in 
fact keep an entirely open mind on the subject, and modify our aproach to 
the extent necessary to understand the data, and others may well do the same.

No one can predict what RHIC will reveal with any credible assurance.  
This paper takes a different approach and therefore will be of vlaue whether 
it fits the data or not.

It should be noted that if multiple bubbles are created in a single
event, each will result in rapidity bumps, strangeness, and anti-baryon-baryon
enhancements etc. in that rapidity interval, which corresponds to that
bubble. Localized QGP bubble formation, probably depends on a first order
transition, however it should also be expected that any unusual
occurrence in a local rapidity region which survives to the end state will 
show up approximately in that rapidity region in the final single event.  
Of course if QGP is made abundantly over large regions of an event our 
bubble techniques will become insensitive.  But it is considered unlikely 
this will occur in all events, and we can study those events where 
plasma regions (or bubbles are localized).

Of course, single events using various selection criteria, can be lumped
together for statistical and overall view and analysis reasons. However 
in doing this great care must be taken to avoid introducing unwanted 
biases in the result.

Finally, these results are to be taken only as an indication of the
promise of the possibility of striking signals of a QGP occurring in single 
events and being detected  in STAR. The correlations of the pions, kaons, 
protons, and anti-protons, and the detailed characteristics of the events 
will of course be important. Our approach and ideas will of course change as 
detailed data is obtained and analyzed. We will keep an open mind as the 
observations come in.

It is important to note that various energies and various beam nuclei from
Au+Au down to p+p will be used and these results will impact on any final 
conclusion. Furthermore it is not necessary to explain every observation. 
Finding striking and unusual events some of the time could  well lead to 
establishment of a QGP if it naturally explains them, and no viable 
alternate explanation is found, and there is no contradiction of
the QGP evidence by other characteristics of the data.

The event generator VNI[29] will also be used in future work. VNI is 
Monte-Carlo event-generator for leptons, hadron, and nucleus collisions 
on each other. It uses the real-time evolution of parton cascades in 
conjunction with a self-consistent hadronization scheme, as well as the 
development of hadron cascades after hadronization. The parton cascading 
in nucleus-nucleus collisions leads to space-time regions of high energy 
density. These regions can become the source of plasma bubbles like the 
ones described above. In the future we plan to use VNI for further work 
on bubbles. We plan to use VNI as a source of bubble production.However 
we must address the problem of how to determine or estimate the energy 
transmitted to the bubble by non-perturbative QCD interactions. VNI will 
place the bubble in a hadronic final state which could give a realistic 
transport of its emitted hadrons to the final detection of plasma signals.

New unexpected phenomena beyond QCD could be observed if they occur at
RHIC, and in that event we plan to use our own and newly developed event
generators to investigate them. It is interesting to note that  a recently 
proposed bubble phenomenon[30,31] as described below can be investigated 
with our techniques and generators.  Any assumptions about STAR performance, 
such as DE/DX, planned time-of-flight performance, etc. are to be considered 
only as our personal estimates for the purpose of comparison with our 
event generator predictions.

\section {Odd CP Bubbles of Metastable Vacuum}

Kharzeev and Pisarski[30] expect that bubbles of metastable vacua with
odd CP will induce a net flow of pion charge.  This flow can be
modeled with a parallel chromo electric and magnetic field.  A quark
traveling in this field will drift up in a spiral, while an anti-quark
will drift down.  These added impulses will end up in positive charged
pions traveling up and negative charged pions traveling down. Kharzeev[31] 
has estimated an average impulse of 30 MeV/c for a quark crossing the 
diameter of the bubble.

If we place the electric $\vec{E}$ and the magnetic $\vec{B}$ field 
along the $x$-axis, then we can write down the added impulse for 
positive pions ($\pi^+$) as vector equations 1:
$$P'_{px} = P_{px} + P_{E}$$
$$P'_{py} = P_{py} - {P_{B}P_{pz}\over P_{pTOT}}\eqno (1)$$
$$P'_{pz} = P_{pz} + {P_{B}P_{py}\over P_{pTOT}},$$
where $\vec{P_p'}$ is the charged $\pi^+$ momentum
$\vec{P_p}$ is the $\pi^+$ momentum without the added
effect and $P_{pTOT}$ is the magnitude of the momentum
$\vec {P_p}$.  $P_{E}$ is the impulse which comes from the 
electric field acting, while $P_B$ comes from the magnetic field.  
Vector equations 2 for negative pions ($\pi^-$) is:
$$P'_{mx} = P_{mx} - P_E$$
$$P'_{my} = P_{my} + {P_B P_{mz}\over P_{mTOT}}\eqno (2)$$
$$P'_{mz} = P_{mz} - {P_B P_{my}\over P_{mTOT}},$$
where $p$ for positive has been replaced by $m$ for minus.  Also the
sign of the impulse has changed.

We will model the added impulse by assuming that we start with a bubble
of radius $r_0$ that has quarks uniformly distributed inside the bubble.

The point $P$ of the quark inside the bubble is given by equation (3):
$$x_p = r_p \sin\theta_p \cos\phi_p$$
$$y_p = r_p \sin\theta_p \sin\phi_p\eqno (3)$$
$$z_p = r_p \cos\theta_p,$$

\noindent where $x_p$, $y_p$, and $z_p$ are the Cartesian coordinates 
and $r_p$, $\theta_p$, and $\phi_p$ are the spherical coordinates.  From 
this point $P$ the quark travels in a line given by its momentum
$$x = {P_x\over P_z} (z-z_p) + x_p$$
$$y = {P_y\over P_z} (z-z_p) + y_p,\eqno (4)$$
where $P_x$, $P_y$, and $P_z$ are the momentum components of the quark.  
The quark will pass through the surface of the bubble when
$$r^2_0 = x^2 + y^2 + z^2.\eqno (5)$$
We now substitute 4 into 5, we have
$$r_0^2 =({P_x\over P_z} (z-z_p) + x_p)^2
+ ({P_y\over P_z} (z-z_p) + y_p)^2 + z^2\eqno (6)$$
Equation (6) can be rewritten in quadratic form using $z'$
$$({P_x^2\over P_z^2} + {P_y^2\over P_z^2} + 1) (z')^2 + 2 ({P_x
x_p\over P_z} + {P_y y_p\over P_z} + z_p) (z') + r^2_p - r^2_o = 0,
\eqno (7)$$ 
where $z' = z  - z_p$.

Using vector notation equation  (7) becomes
$${P_p^2\over P_z^2} (z')^2 + {2\over P_z} (\vec P_p \cdot \vec r_p)
(z') - (r^2_o - r^2_p) = 0.\eqno (8)$$

Thus we have two solutions:
$$ z' = z-z_p = {(\vec P_p \cdot \vec r_p) \pm \sqrt{(\vec P_p \cdot
\vec r_p)^2 + P^2_p (r^2_o - r^2_p)}\over P^2_p}, \eqno (9)$$
because a line strikes a sphere twice when passing through it.

The distance the quark travels is given by
$$dist = \sqrt{(x-x_p)^2 + (y-y_p)^2 + (z-z_p)^2}.\eqno (10)$$

Using equation (4), the distance equation (10) becomes:

$$dist = \sqrt {{P_x^2\over P_z^2} + {P^2_y\over P_z^2} + {P_z^2\over
P_z^2}}\;\;\; {\left | (z-z_p)\right |} \eqno (11)$$

\noindent or
$$dist = {P_p\over \vert P_z\vert} \vert (z-z_p)\vert$$
\noindent Plugging in (9) into (11), we obtain:
$$dist = \left | {(\vec P_p \cdot \vec r_p)\pm\sqrt{\vec P_p\cdot \vec r_p +
P^2_p (r^2_o - r^2_p)}\over P_p}\right | \eqno (12)$$
Spherical coordinate system (3) makes equation (12) look very simple

$$dist = r_p(\sin\theta\sin\theta_p(\cos\phi\cos\phi_p +
\sin\phi\sin\phi_p) + \cos\theta\cos\theta_p) $$ 
$$\pm\;\sqrt{r^2_p(\sin\theta\sin\theta_p(\cos\phi\cos\phi_p + 
\sin\phi\sin\phi_p)+\cos\theta\cos\theta_p)^2+r^2_0-r^2_p}\eqno (13)$$

where $\theta$ and $\phi$ are the angles of the momentum vector $\vec
P_p$, $\theta_p$ and $\phi_p$ are the angles of the point $P$ with $r_p$ 
being the radius of the point and $r_0$ the radius of the bubble.

We now want to obtain the distribution of $dist$ taking both solutions for
every point $P$, when the points are uniformly placed inside the bubble.
The easiest way to do this calculation is using Monte Carlo methods.  The
result of such a numerical method leads to

$${dN\over d_{dist}} \;\propto\; \cos ({\pi dist\over 4 r_o}).\eqno (14)$$

The impulse that the quark gets is directly proportional to the
distance.  Since 2 $r_o$ is the maximum distance, we can set the maximum
impulse $(\Delta P_{max})$ equal to this value and easily write down the
impulse distribution.
$${dN\over d\Delta P} \;\propto\; \cos ({\pi\Delta P\over 2\Delta
P_{max}}).\eqno (15)$$
In equations (1) and (2) the value of $P_E$ and $P_B$ can be chosen by
picking a random number $R$ between zero and one and calculating
$$\Delta P = \Delta P_{max} = {2\Delta P_{max}\over \pi} \cos^{-1}
(R).\eqno (16)$$

The next aspect in bubble simulation is the kinematics of the pions
produced by the bubble.  Gyulassy[32] suggested jet fragmentation would
show these added impulse to their pions.  Studies of changes of jet
fragmentation in heavy ion events is already underway[33].  For this
simulation we consider bubble kinematics that would arise from a chiral
low temperature bubble or a bubble that produced pions which look very
much like the regular pions of a heavy ion event.  A set of programs 
using the event generator RHIC::PLASMA is used to simulate bubbles that
have imparted impulses given by equation (16).

Chiral bubbles are spherically symmetric distributions of pions having a
Boltzman energy spectrum of 70 meV temperature.  The chromo $E$ and $B$
field is randomly assumed in any direction.  The E field gives a impulse
to $\pi^+$ along its direction, while the $\pi^-$ impulse is opposite to
its direction.  The impulse of the B field is at right angles to the
direction, where the example of B and E along the X-axis is given by
equations (1) and (2).

Besides a low temperature chiral bubble, let us consider a higher
temperature bubble which expands in the longitudinal beam direction.
The net result leads to a bubble which blends into the regular
background of RHIC::EVENT.

The program generates a bubble by a regional tagging scheme which 
takes particles from the initial RHIC::EVENT collision and converts 
the energy into a bubble.  The tagging region is defined as a circular 
cross sectional area located within the center of the overlap region
with radius $R_{tag}$.  The bubble has an initial energy, where its rest 
frame is given by the total momentum energy four vector of the sum of 
the tagged particles that are used to create the bubble.

\section {Summary and Conclusions}

In this paper we have discussed a possible way striking signals for a
QGP could be generated at RHIC and detected. We have extended and
generalized the original scenario proposed by Van Hove[6,7], by 
developing a detailed model with associated event generators to explore 
this scenario.  Some of the results have been presented here.

If QGP bubbles or localized rapidity regions of QGP are created in some
of the RHIC collisions (probably most likely for highest energy Au on Au),
and the characteristics of their hadron emissions preserved, at least
substantially in part of the events final state, we have shown that 
reasonable event generators that we developed can provide striking 
signals in individual events, even under the modest assumption that only 
4.5\% of the available energy (corresponding to the available energy  
for plasma formation generated in a central collision contained within 
a radius of about 1.3~$fm$) forms the QGP bubble. If larger amounts of
the available energy form a QGP bubble (or a localized rapidity region)
the signals will scale with the QGP energy, and could become truly 
dramatic (see Figs. 9-13).

We considered how the STAR detector, for example,  might observe some 
of  these modest 4.5\% of the available energy QGP signals (figs. 5-8). 
If we assume $\approx$15\% of the available energy is converted to 
plasma, we obtain truly dramatic signals (see figs. 9-13) and indicate 
how the STAR detector might observe some of these.

No one can predict what RHIC will reveal with any credible assurances.  
However when data is obtained with RHIC we will generate what is needed
and relevant to confront our model, and in fact keep an entirely open mind 
on the subject, and modify our approach to the extent necessary to 
understand the data, and others may well do the same.

This paper and the event generators we have developed take a different 
approach, and therefore will be of value whether it can explain 
characteristics of the data or not.

Finally it should be noted that our RHIC::EVENT and RHIC::PLASMA
generators can be used to investigate ``odd CP bubbles'' and likely can 
be adapted to many new investigations. We also discussed how the VNI 
event generator could be applied to QGP formation.

\begin{figure}
\begin{center}
\mbox{
   \epsfysize 5.2in
   \epsfbox{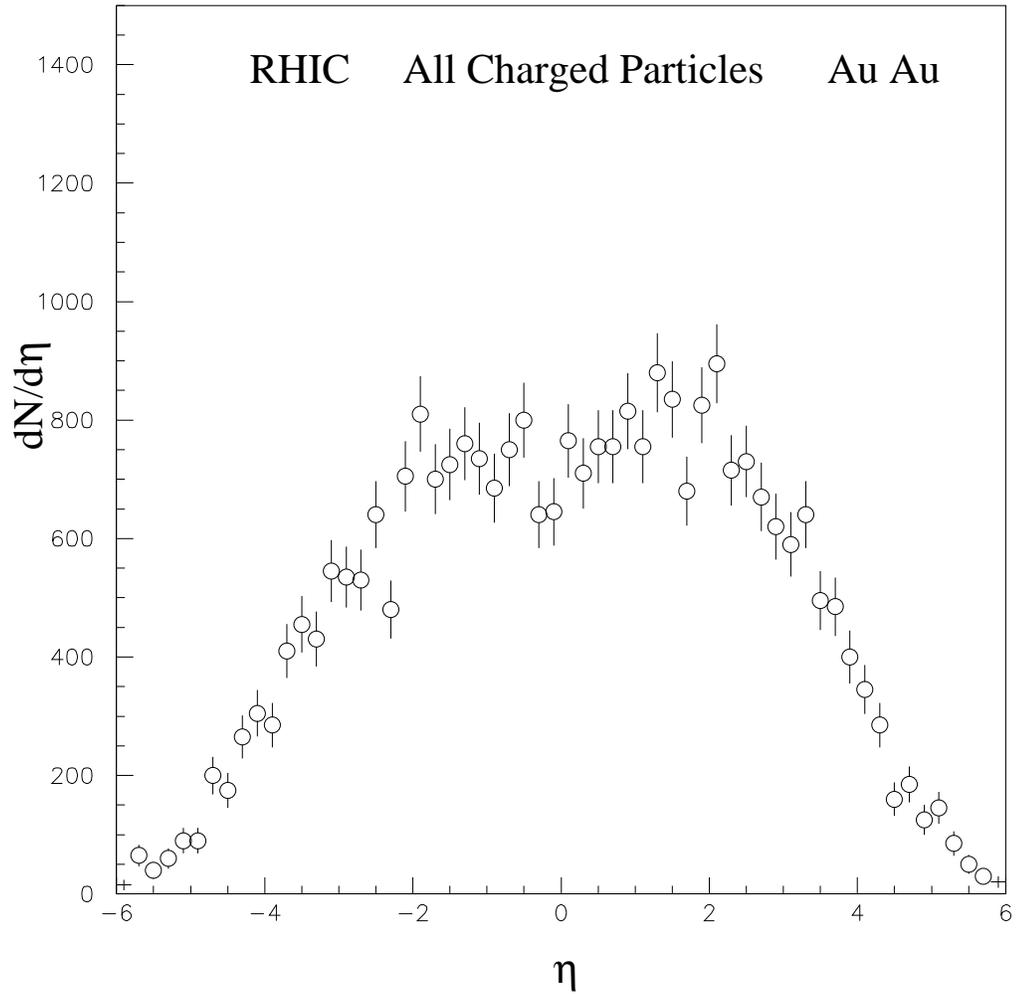}}
\end{center}
\vspace{10pt}
\caption{The  generated pseudorapidity distribution of all charged 
particles from a single ``central'' 100 Gev/n  AU+AU  RHIC::EVENT The 
expected central plateau without any evidence of structure is observed.}
\end{figure}

\begin{figure}
\begin{center}
\mbox{
   \epsfysize 5.2in
   \epsfbox{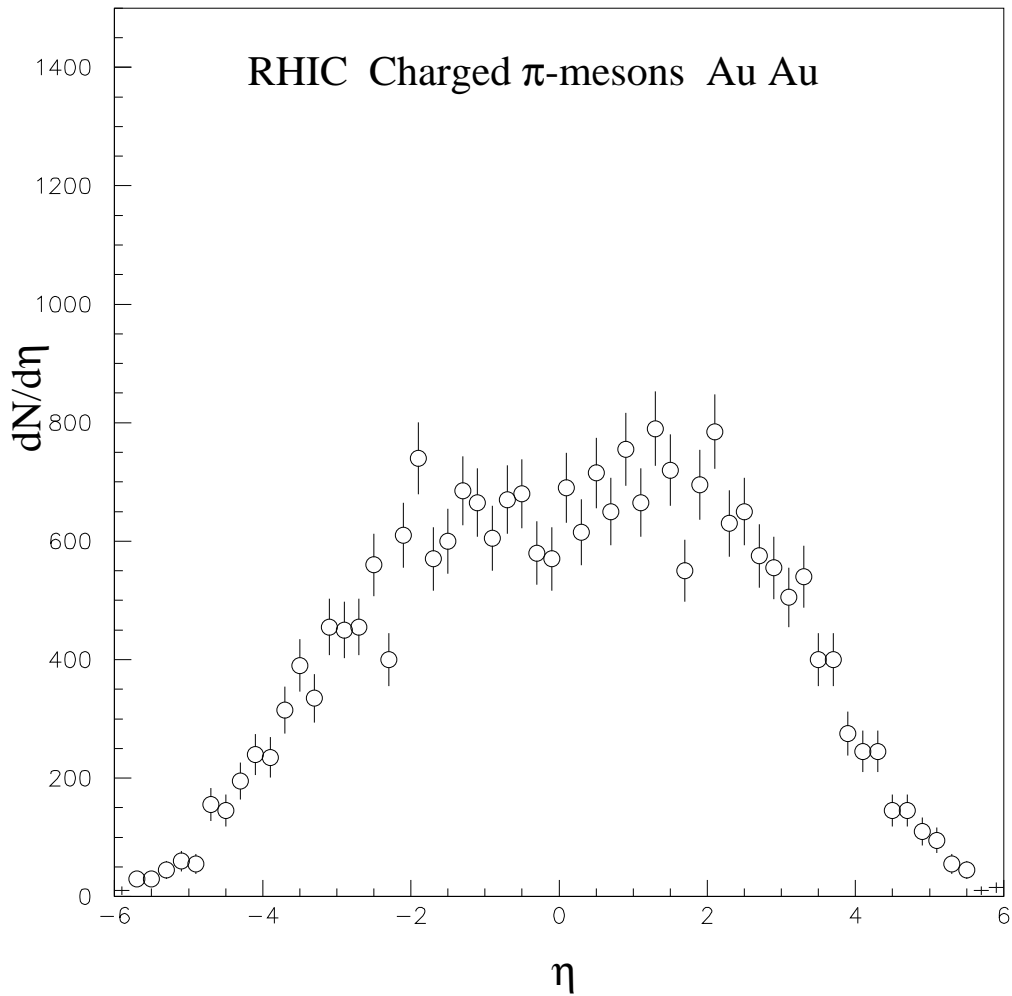}}
\end{center}
\vspace{10pt}
\caption{The generated pseudorapidity distribution of all charged  
pions from a single ``central'' 100 Gev/n  AU+AU  RHIC::EVENT. The 
expected central plateau with no evidence of structure is observed.
Negative pions can be identified approximately by assuming all negative 
particles are pions.  If one removes kaons and protons (or anti-protons)
from a momentum region, what remains is almost entirely pions.}
\end{figure}

\begin{figure}
\begin{center}
\mbox{
   \epsfysize 5.2in
   \epsfbox{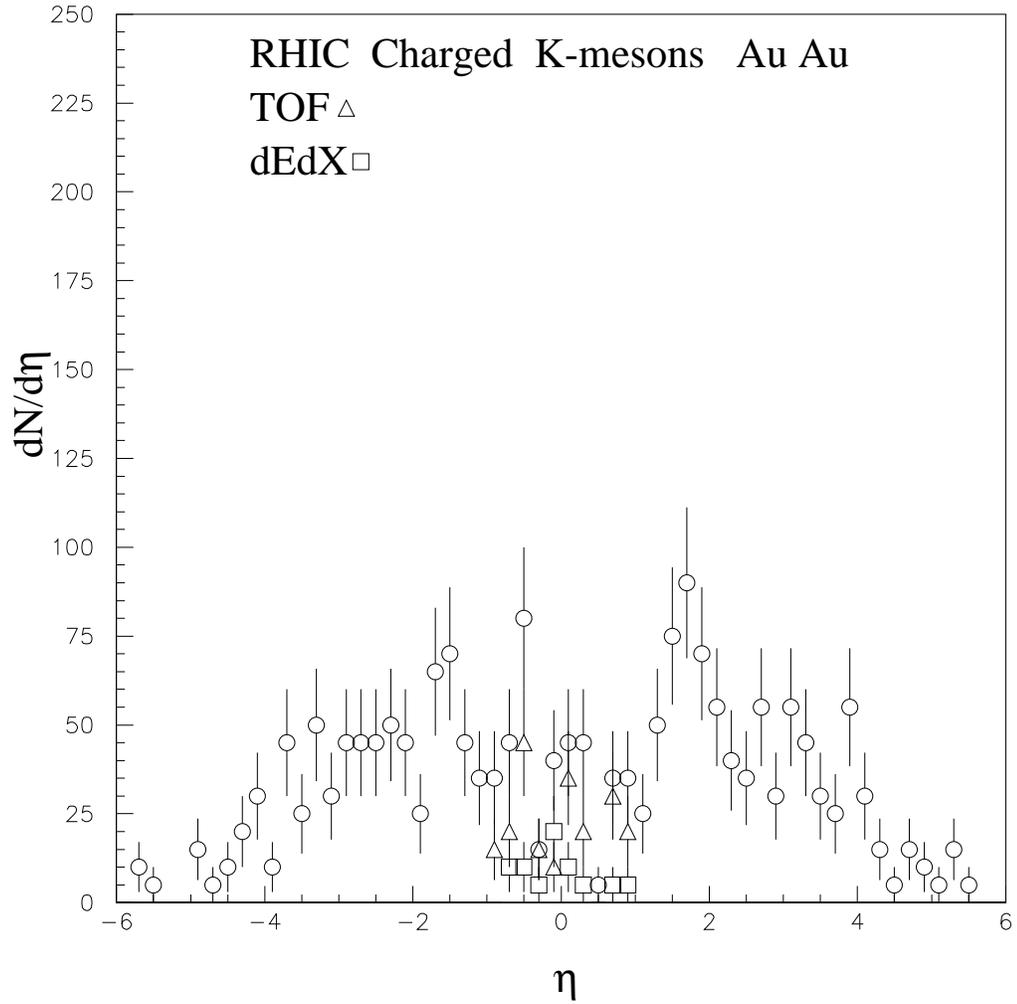}}
\end{center}
\vspace{10pt}
\caption{The generated pseudorapidity distribution for charged K-mesons 
from a single ``central'' AU+AU RHIC::EVENT. The STAR identified K-mesons 
estimated by DE/DX (ionization) and TOF (Time-of-Flight) with the 
geometric efficiency only for the STAR detector central TPC. A central 
plateau with no evidence for structure is observed.}
\end{figure}

\begin{figure}
\begin{center}
\mbox{
   \epsfysize 5.2in
   \epsfbox{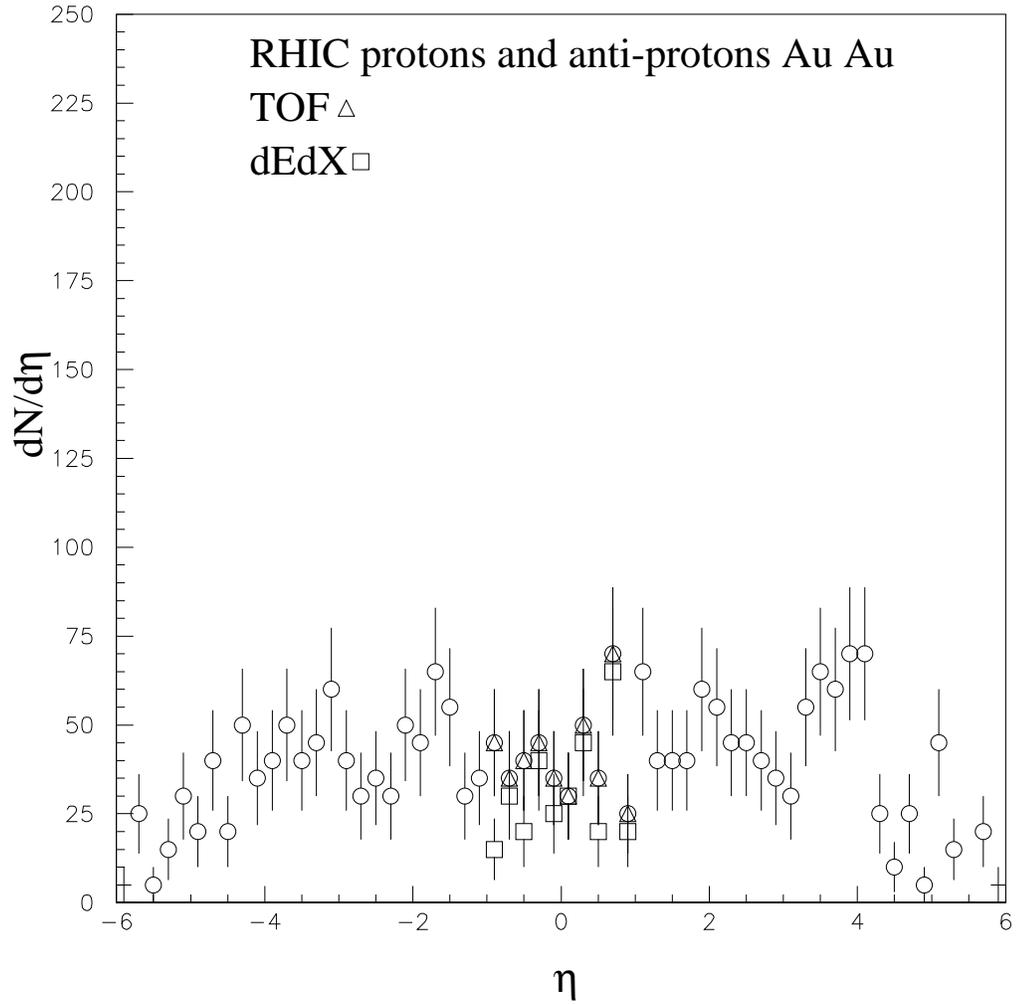}}
\end{center}
\vspace{10pt}
\caption{The generated pseudorapidity distribution for protons and 
antiprotons from a single ``central'' AU+AU RHIC::EVENT with STAR 
DE/DX (ionization) and TOF (Time-of-Flight) identification with the 
geometric efficiency only for the central TPC shown. 
A central plateau with no evidence of structure is observed.}
\end{figure}

\begin{figure}
\begin{center}
\mbox{
   \epsfysize 5.2in
   \epsfbox{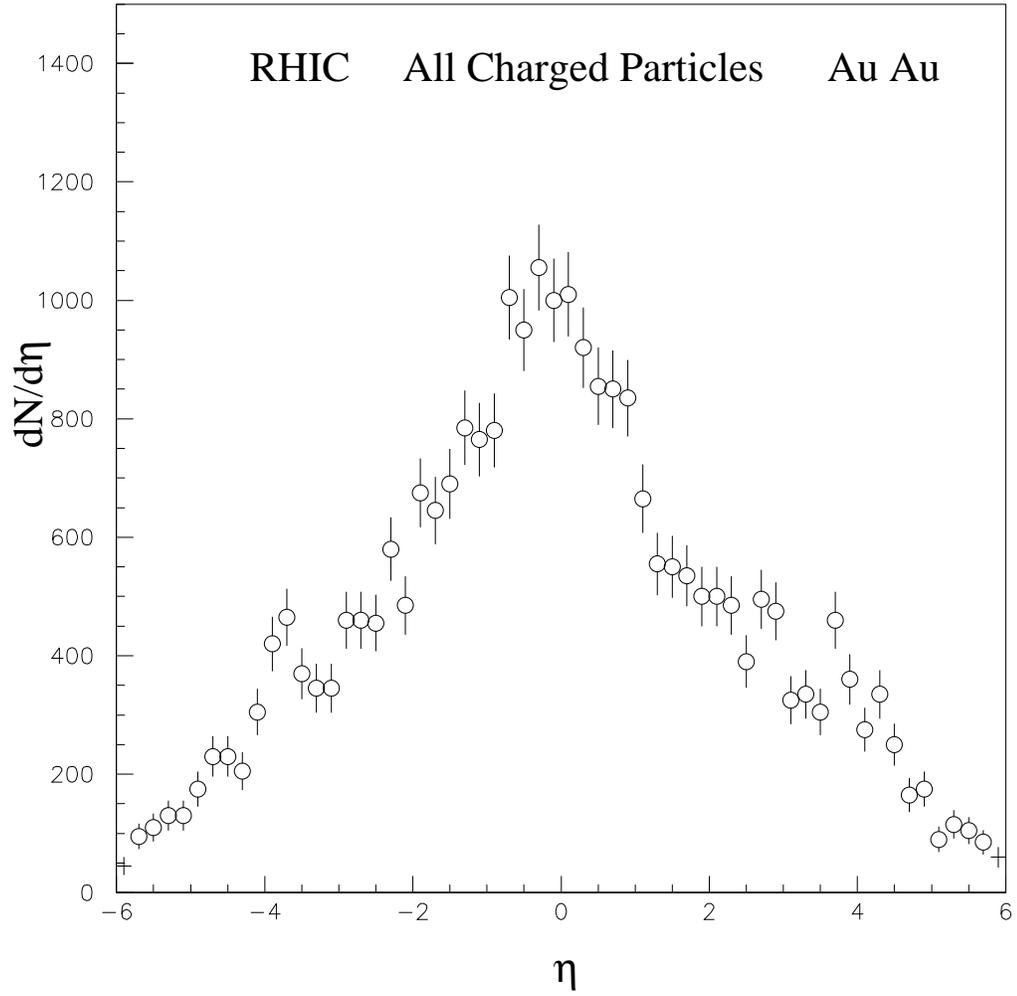}}
\end{center}
\vspace{10pt}
\caption{The generated pseudorapidity distribution of all charged 
particles from a single RHIC::PLASMA ``central'' EVENT. Note the 
well-defined $\eta$ central peak in contrast to the structureless 
result in Fig. 1 for RHIC::EVENT.  The bubble energy was 4.5\% of the
available energy.}
\end{figure}

\begin{figure}
\begin{center}
\mbox{
   \epsfysize 5.2in
   \epsfbox{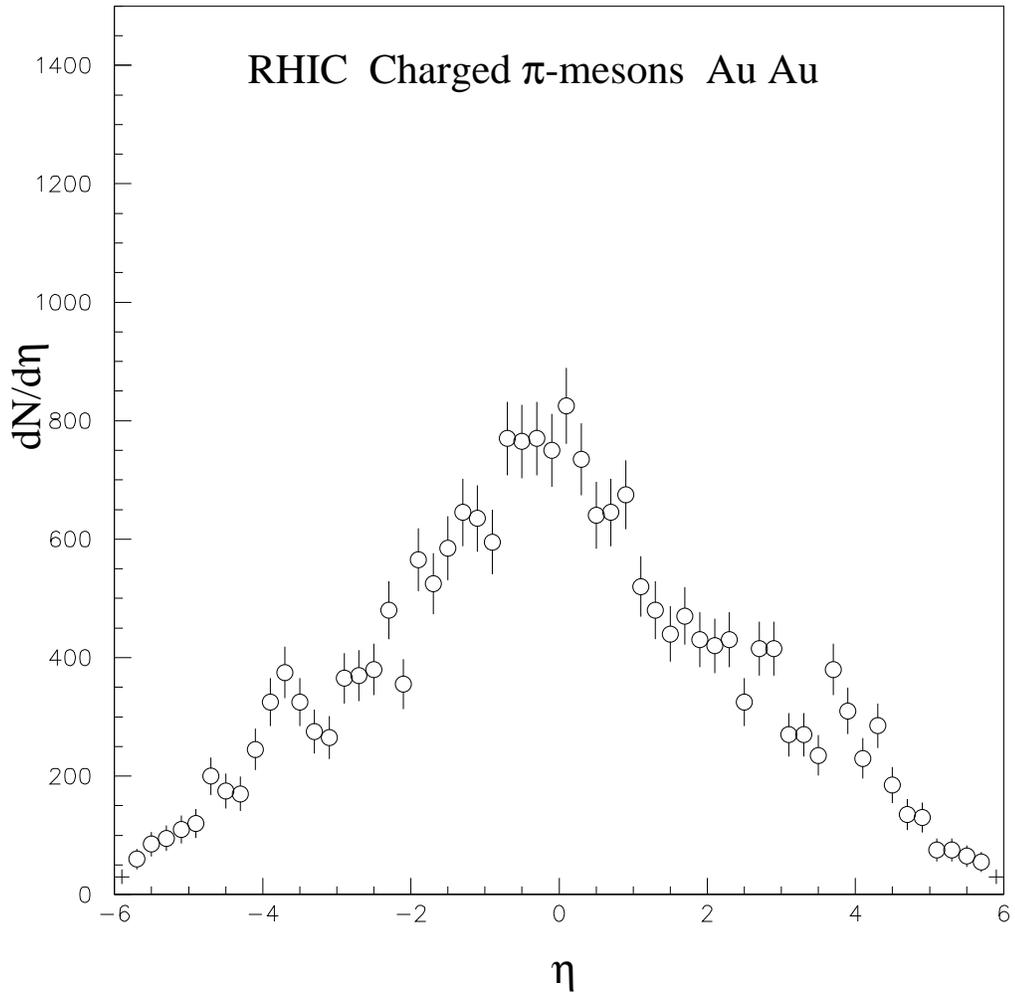}}
\end{center}
\vspace{10pt}
\caption{The generated pseudorapidity distribution for all charged pions 
from a single ``central'' 100 Gev/n AU+AU RHIC::EVENT again showing 
a well defined central bubble peak.  Negative pions can be identified 
approximately by assuming all negative paticles are pions.  If one 
removes kaons and protons (or anti-protons from a momentum region
what remains is almost entirely pions. The bubble energy was 4.5\% of the
available energy.}
\end{figure}

\begin{figure}
\begin{center}
\mbox{
   \epsfysize 5.2in
   \epsfbox{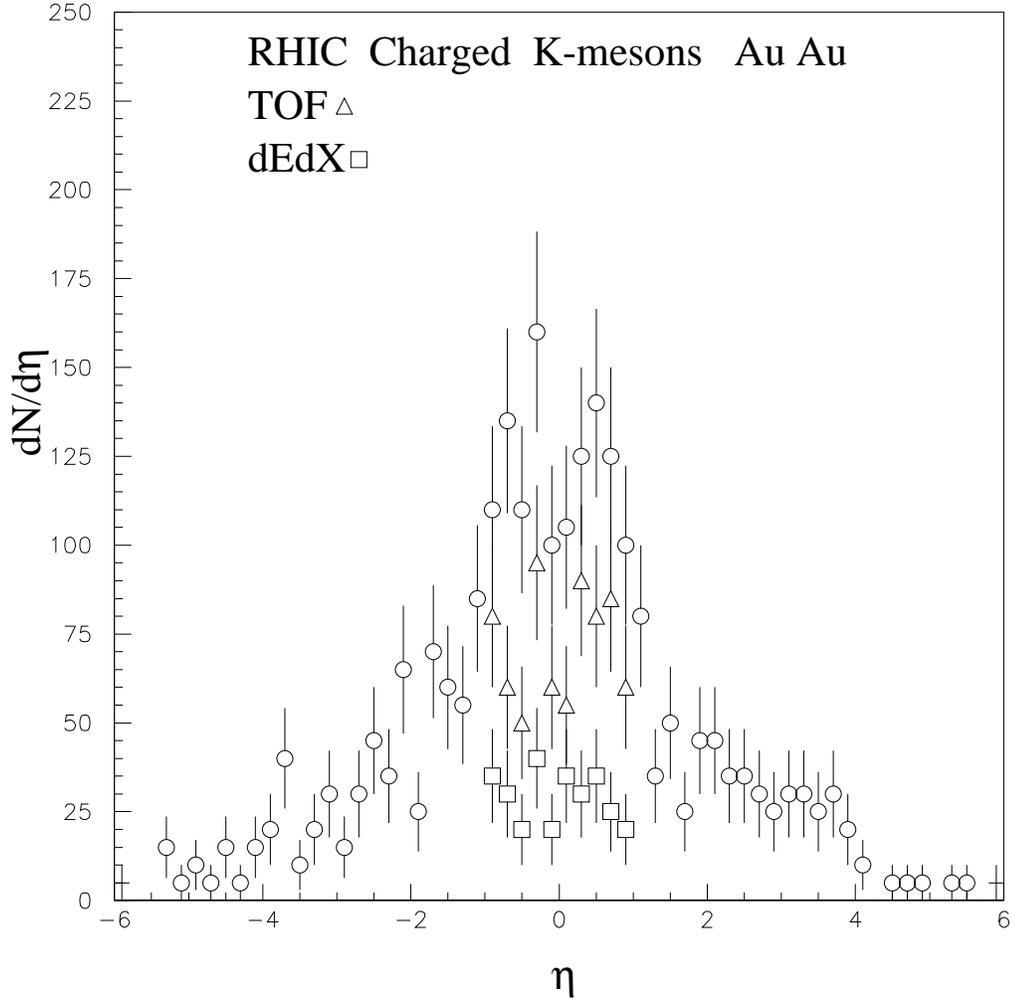}}
\end{center}
\vspace{10pt}
\caption{The generated pseudorapidity distribution for all charged 
K-mesons from a single ``central'' 100 Gev/n AU+AU RHIC::EVENT showing 
a peak and that TOF can be effective especially if larger than 4.5\% of 
energy is converted to plasma in some events. The geometric efficiency
only for the central TPC was used. The bubble energy was 4.5\% of the
available energy. }
\end{figure}

\begin{figure}
\begin{center}
\mbox{
   \epsfysize 5.2in
   \epsfbox{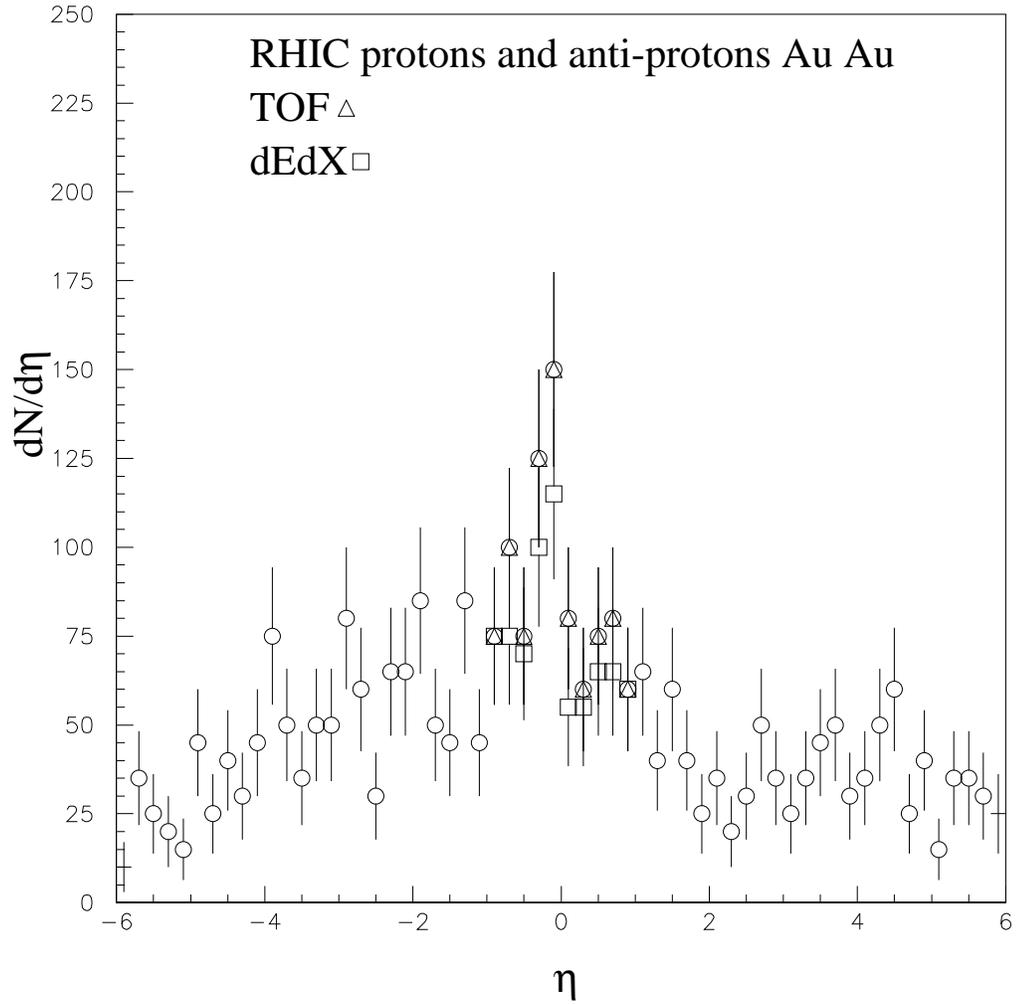}}
\end{center}
\vspace{10pt}
\caption{The generated pseudorapidity distribution for protons and 
anti-protons, the geometric efficiency only for the central TPC was
used showing a peak which is well identified by TOF and  to some extent
indicated by DE/DX. The bubble energy was 4.5\% of the available energy.}
\end{figure}

\begin{figure}
\begin{center}
\mbox{
   \epsfysize 5.2in
   \epsfbox{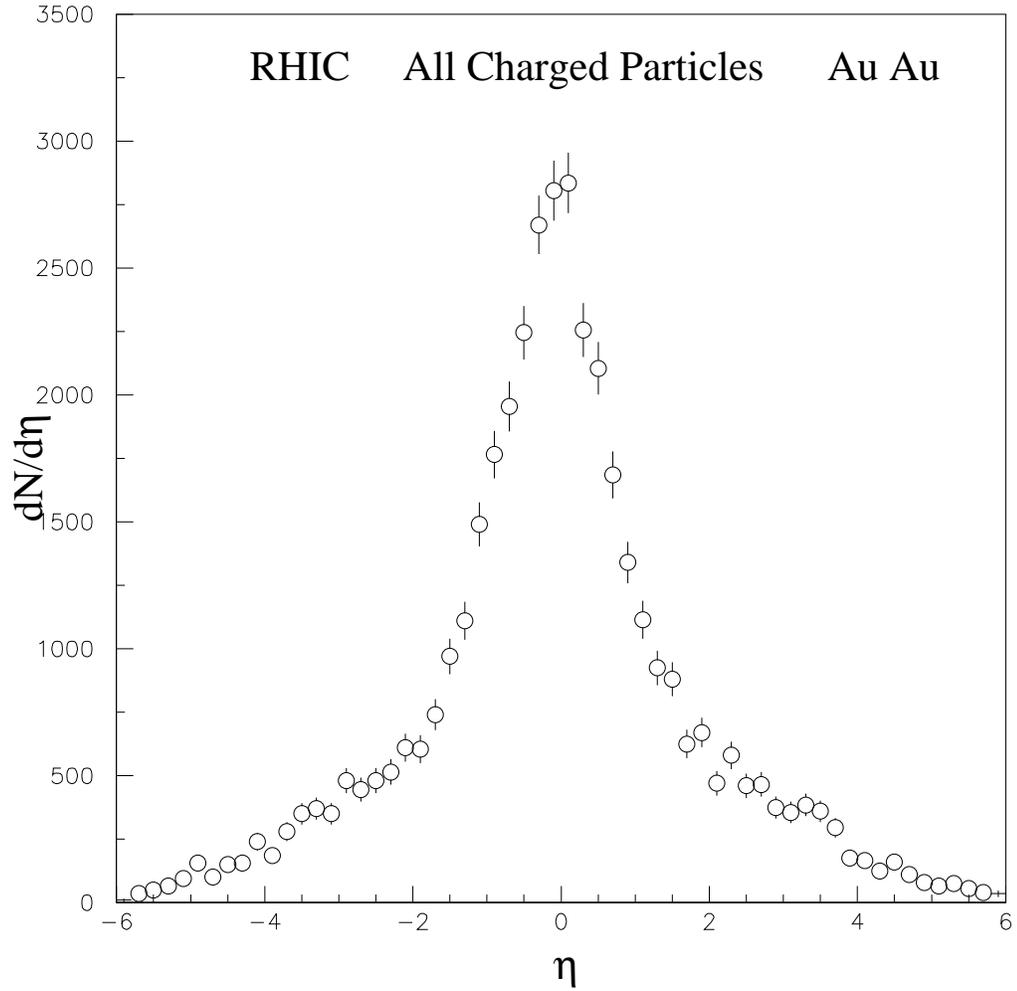}}
\end{center}
\vspace{10pt}
\caption{The generated pseudorapidity distribution of all charged particles 
from a single ``central'' 100 Gev/n AU+AU RHIC::PLASMA event.  The height 
of the bubble peak is about four times the background level. The bubble 
energy was $\approx$ 15\% of the available energy. }
\end{figure}

\begin{figure}
\begin{center}
\mbox{
   \epsfysize 5.2in
   \epsfbox{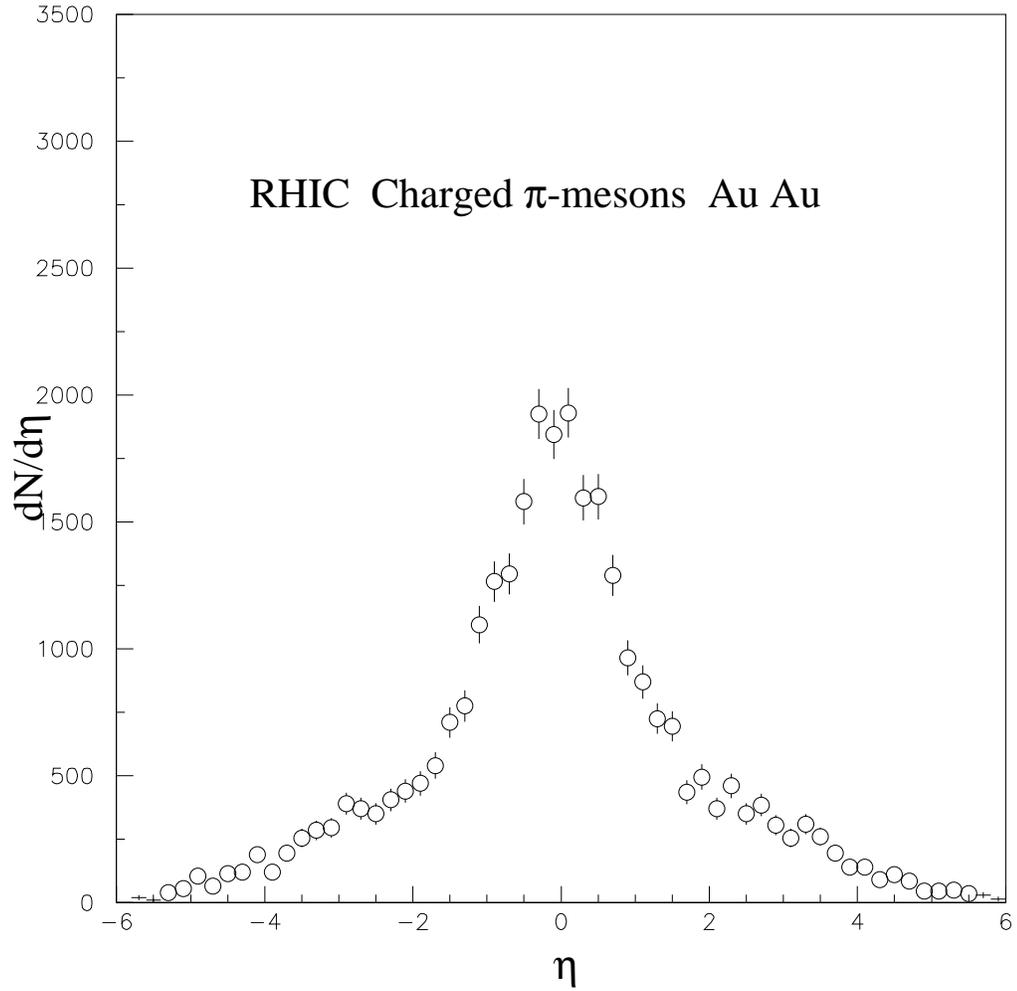}}
\end{center}
\vspace{10pt}
\caption{The generated pseudorapidity distribution of all charged pion 
from a single ``central'' 100 Gev/n AU+AU  event.  The height 
of the bubble peak is about 3.5  times the background level. The bubble 
energy was $\approx$ 15\% of the available energy. }
\end{figure}

\begin{figure}
\begin{center}
\mbox{
   \epsfysize 5.2in
   \epsfbox{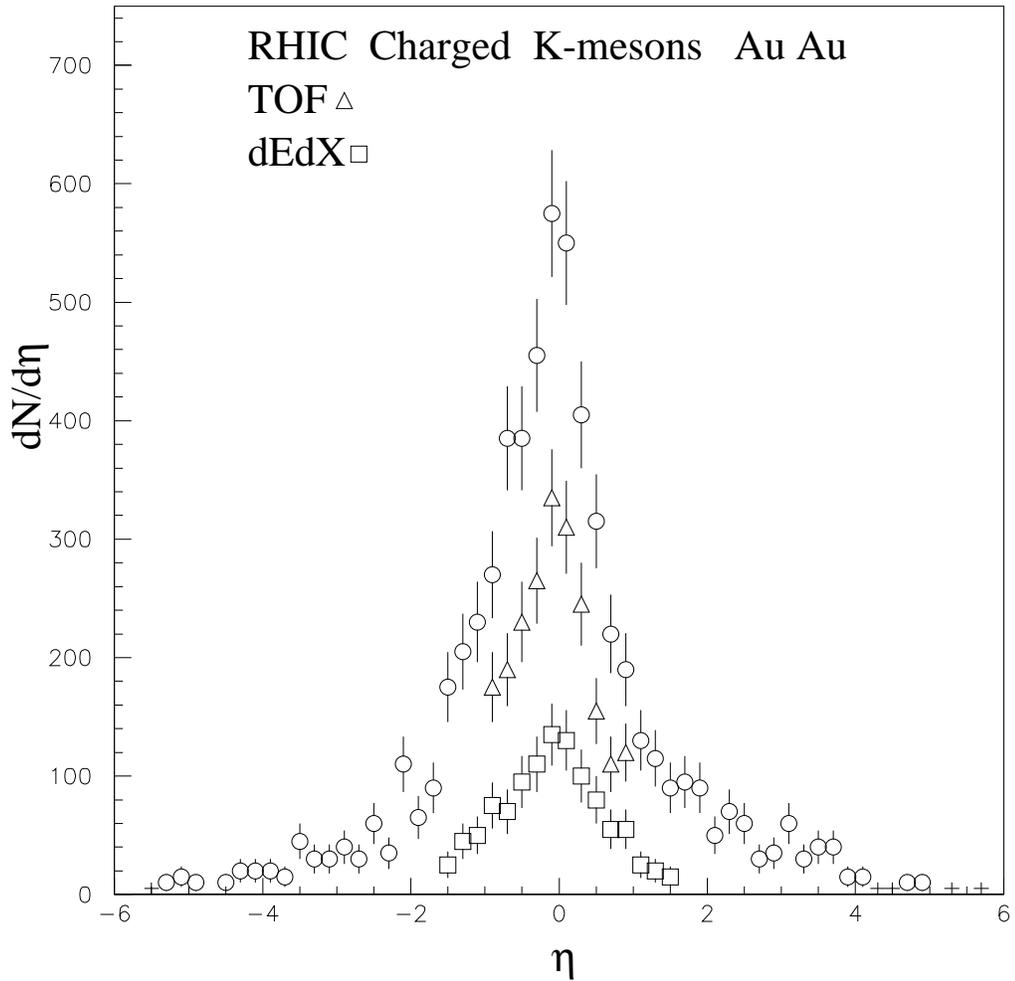}}
\end{center}
\vspace{10pt}
\caption{The pseudorapidity distribution for all charged K-mesons 
from a single ``central'' 100 Gev/n event, showing a peak about four 
times the background level.  Both the TOF and DE/EX particle 
identifications show peaks 2-2.5 times the background level, but the 
TOF has much more statistics and detects more energetic particles.
The bubble energy was $\approx$ 15\% of the available energy.  }
\end{figure}

\begin{figure}
\begin{center}
\mbox{
   \epsfysize 5.2in
   \epsfbox{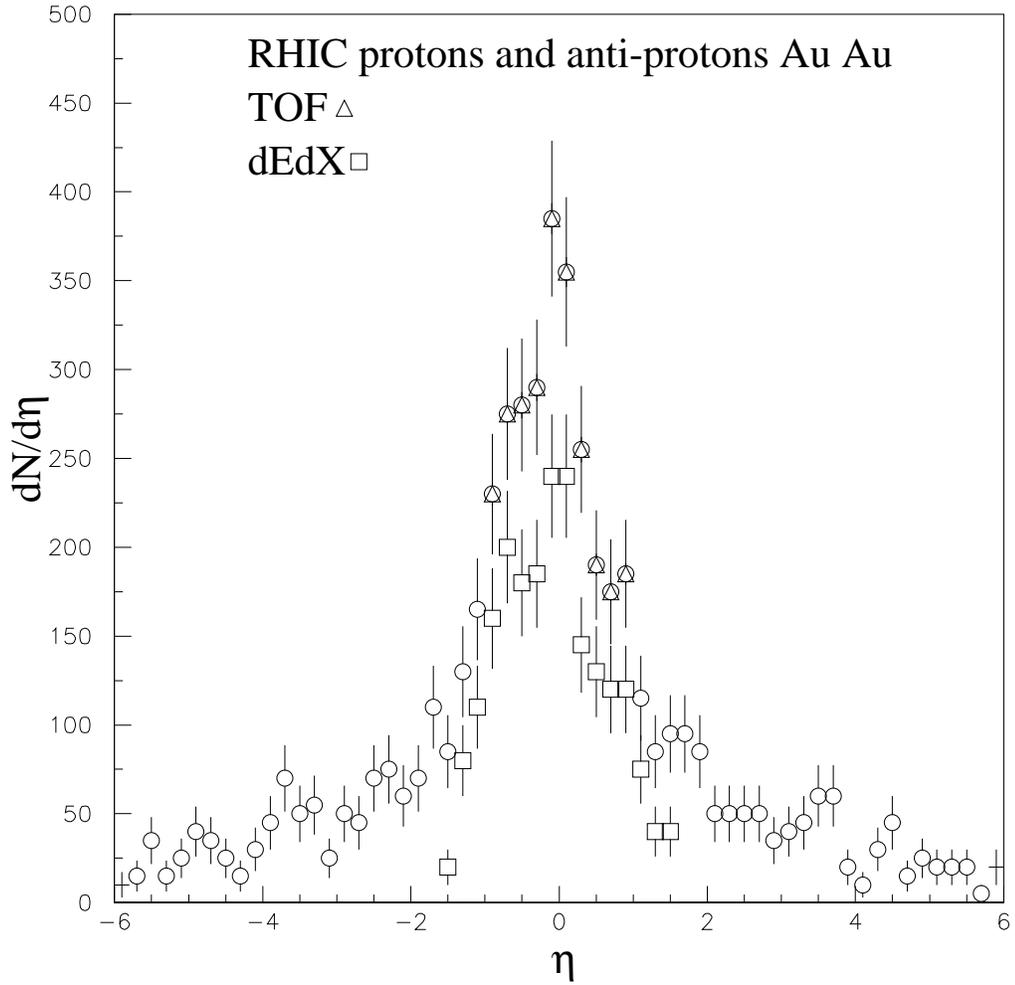}}
\end{center}
\vspace{10pt}
\caption{The pseudorapidity distribution for all protons and anti
protons from a single ``central'' 100 Gev/n event, showing a peak about 
four times the background level, but the TOF has much more statistics 
and detects more energetic particles. The bubble energy was $\approx$ 
15\% of the available energy. }
\end{figure}

\begin{figure}
\begin{center}
\mbox{
   \epsfysize 5.2in
   \epsfbox{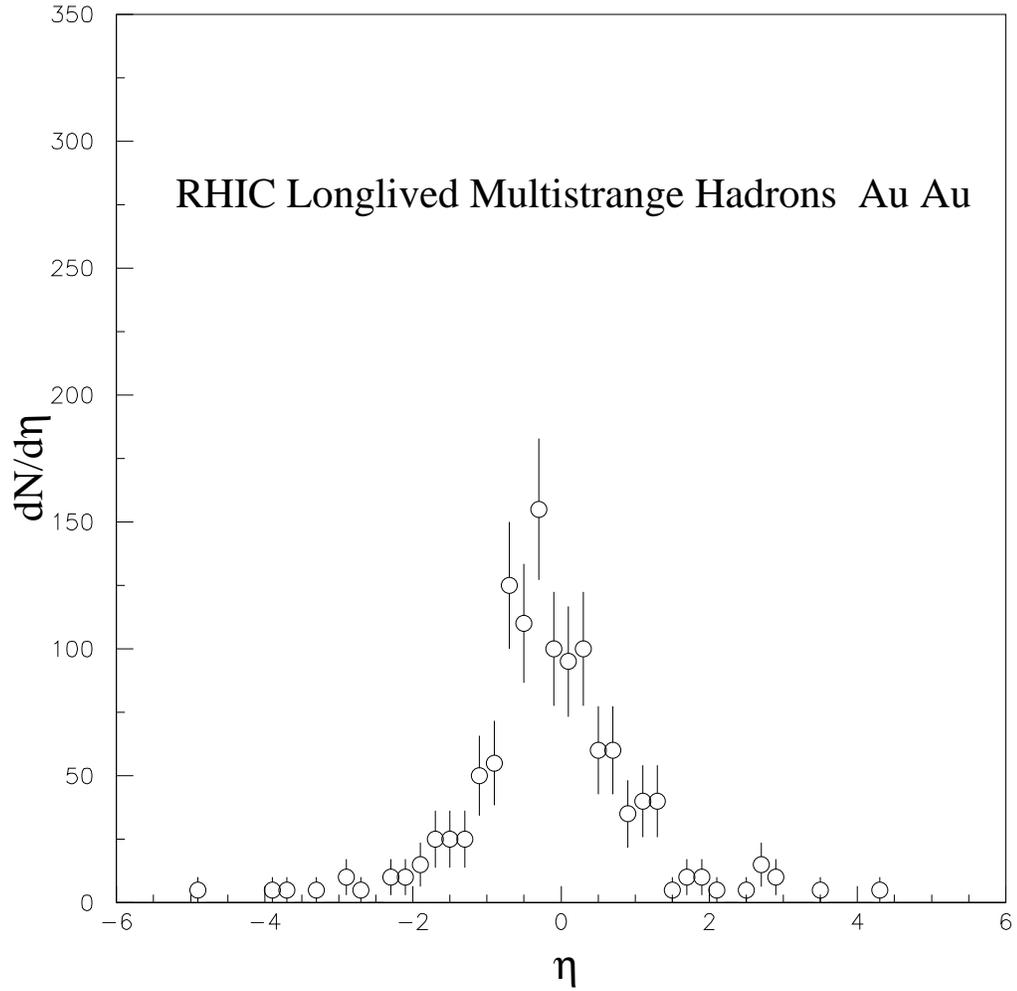}}
\end{center}
\vspace{10pt}
\caption{The pseudorapidity distribution for all long-lived Multistrange
Hadrons ($\Xi$, $\Omega^-$) from a single ``central'' 100 Gev/n event, 
showing a peak level about three times the background level. The bubble 
energy was $\approx$ 15\% of the available energy.}
\end{figure}

\begin{figure}
\begin{center}
\mbox{
   \epsfysize 5.2in
   \epsfbox{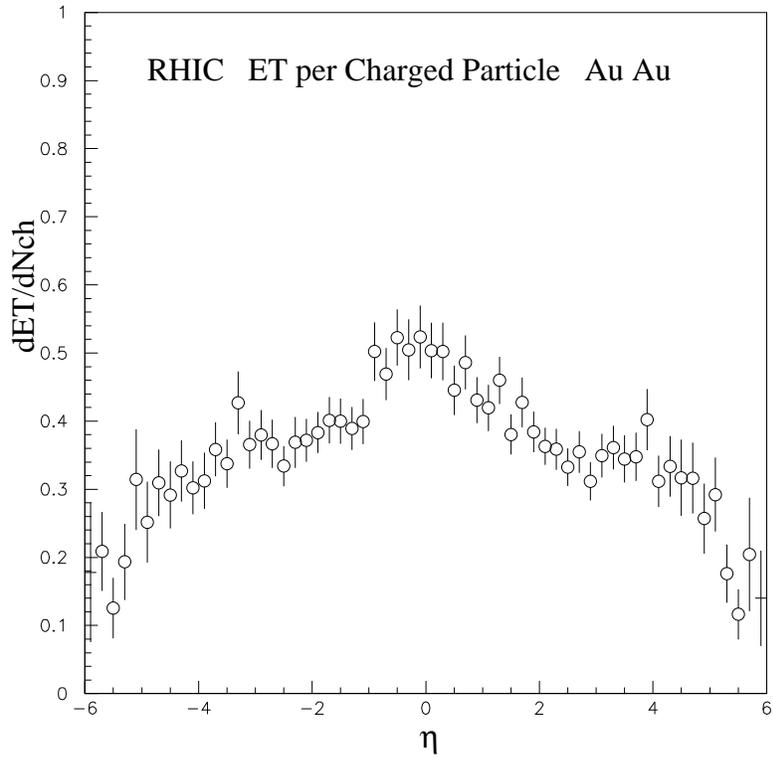}}
\end{center}
\vspace{10pt}
\caption{The pseudorapidity distribution of dET/dNch (average transverse 
energy/per charged particle) for all charged particles from a single 
``central'' 100 Gev/n Au + Au RHIC::EVENT (see Fig. 1).}
\end{figure}
\begin{figure}
\begin{center}
\mbox{
   \epsfysize 5.2in
   \epsfbox{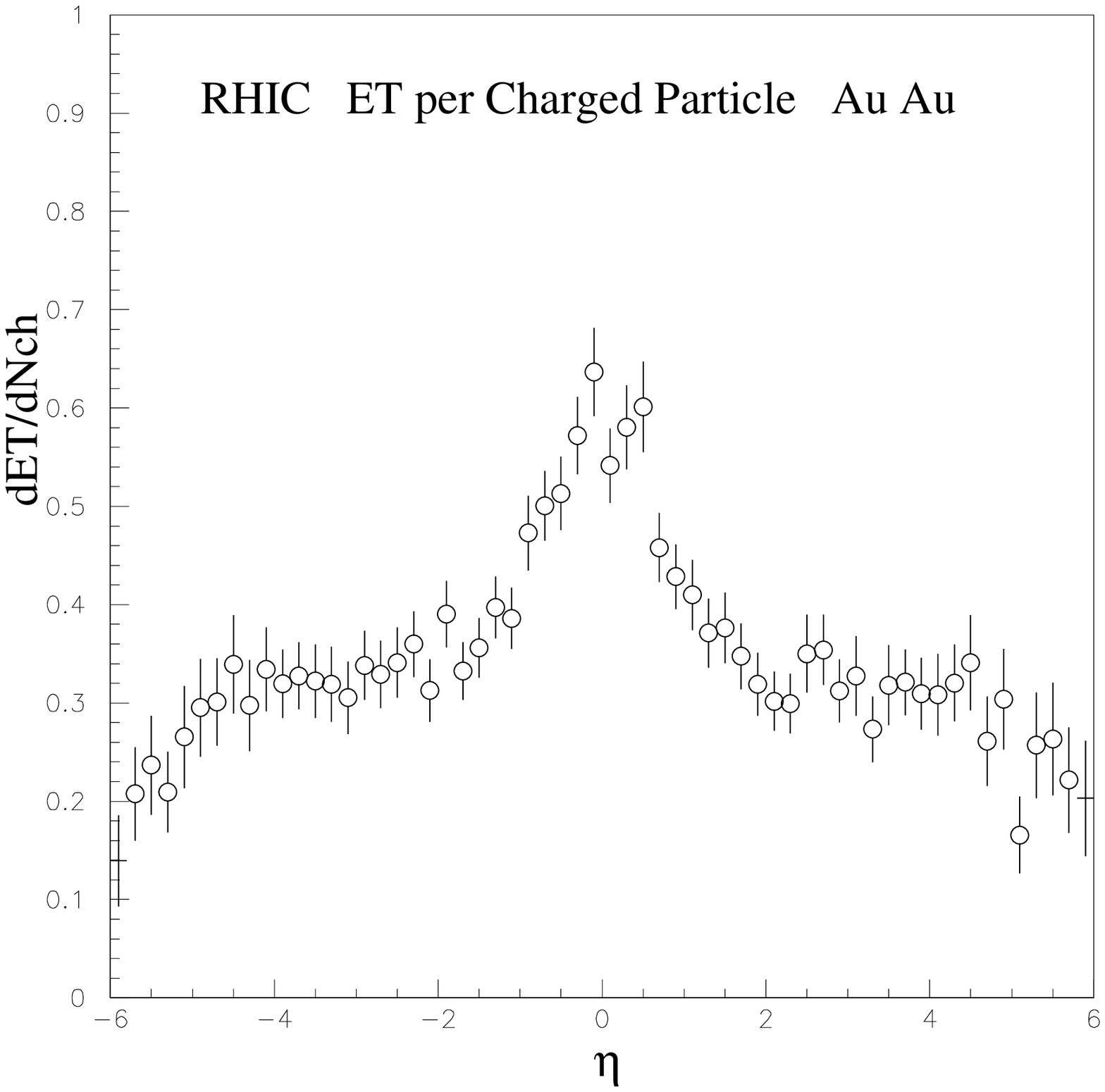}}
\end{center}
\vspace{10pt}
\caption{The pseudorapidity distribution of dET/dNch (average transverse 
energy/per charged particle) for all charged particles from a single 
``central'' 100 Gev/n Au + Au RHIC::PLASMA event (see Fig. 5). The bubble
energy was 4.5\% of the available energy.}
\end{figure}

\begin{figure}
\begin{center}
\mbox{
   \epsfysize 5.2in
   \epsfbox{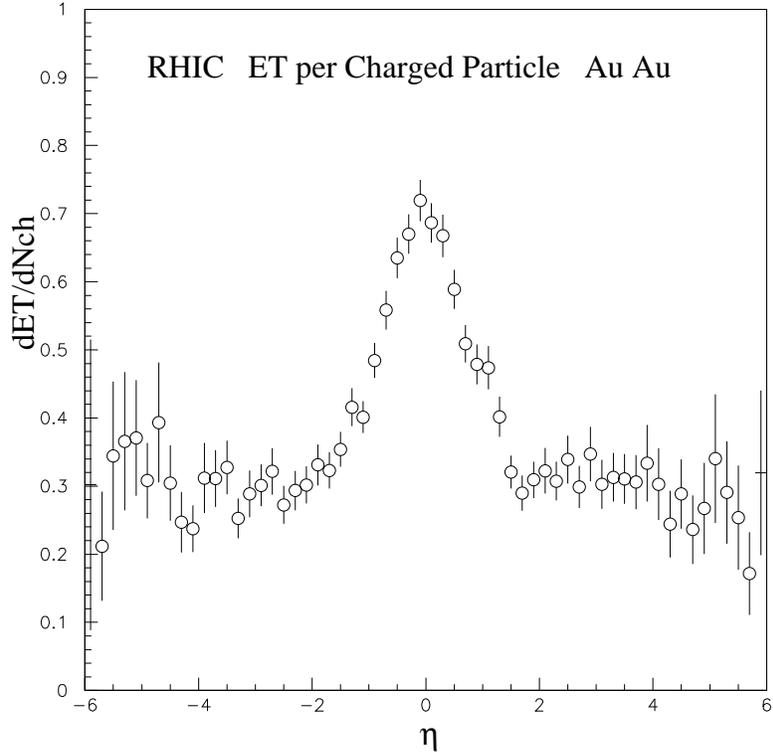}}
\end{center}
\vspace{10pt}
\caption{The pseudorapidity distribution of dET/dNch (average transverse 
energy/per charged particle) for all charged particles from a single 
``central'' 100 Gev/n Au + Au RHIC::PLASMA event (see Fig. 9). The bubble
energy was $\approx$ 15\% of the available energy.}
\end{figure}

\newpage
\centerline {REFERENCES}

[1] Quark Matter Formation and Heavy Ion Collisions: Proc. of the Biefeld
Workshop, May 82.  See papers by Jacobs and Satz, Baym, Kajantie,
McLerran, Gyulassy, Knoll, Kapusta and others.

[2] Quark Matter conferences 1983, BNL; 1984 Helsinki, 1986 Asilomar,
1987 Nordkirschen, 1988 Lenox; 1990 Menton; 1991 Gatlinburg; 1993 Borlange,
1995 Monterey, 1996 Heidelberg, 1997 Tsukuba, 1998 Torino.

[3]  Samios, N.P., RHIC Conceptual Design Report, (Brookhaven National
Laboratory, BNL52195, (1989).

[4]  L. Schroeder and S.J. Lindenbaum, Large Magnetic Spectrometers.
Part II by S.J. Lindenbaum, RHIC Workshop: Experiments for a Relativistic 
Heavy Ion Collider, April 15-19, 1985, P.E. Haustein and C.L. Woody, 
Editors, pp. 227-252 (Brookhaven National Laboratory, Upton, New York, 
1985), BNL51921.

[5] a) L. Schroeder and S.J. Lindenbaum,  Large Magnetic Spectrometer.
Lindenbaum, S.J., part II pp. 227-252 RHIC Workshop: Experiments for a
Relativistic Heavy Ion Collider, April 15-19, 1985, P.E. Haustein and
C.L. Woody, Editors, pp. 211-252 (Brookhaven National Laboratory, Upton, 
New York, 1985).  (b) Lindenbaum, S.J.  An Approximately 4p Tracking 
Magnetic Spectrometer for RHIC.  Proc. of the Second Workshop on 
Experiments and Detectors for a Relativistic Heavy Ion Collider (RHIC), 
Lawrence Berkeley Laboratory, Berkeley, California, May 25-29, 1987, 
Editors, Hans Georg Ritter and Asher Shor, pp. 146-165 (Lawrence Berkeley 
Laboratory, 1988).  (c) Lindenbaum, S.J.  A 4$\pi$ Tracking Magnetic 
Spectrometer for RHIC.  Proc. of the Third Workshop on
Experiments and Detectors for a Relativistic Heavy Ion Collider (RHIC),
Brookhaven National Laboratory, July 11-22, 1988, B. Shivakumar and P.
Vincent, Editors, pp. 82-96  (Brookhaven National Laboratory, BNL 52185).  
(d) Lindenbaum, S.J. (Experimental Collaboration: G. Danby, S.E. Eiseman, 
A. Etkin, K.J. Foley, R.W. Hackenburg, R.S. Longacre, W.A. Love, T.W. 
Morris, E.D. Platner, A.C. Saulys, J.H. Van Dijk, S.J. Lindenbaum, C.S. 
Chan, M.A. Kramer, K. Zhao, N. Biswas, P. Kenney, J. Piekarz, D.L. Adams, 
S. Ahmad, B.E. Bonner, J.A.Buchanan, C.N. Chiou, J.M. Clement, M.D. 
Corcoran, T. Empl, H.E. Miettinen, G.S. Mutchler, J.B. Roberts, J. Skeens) 
A 4p Tracking TPC Magnetic Spectrometer for RHIC.  Proc. of the Fourth 
Workshop on Experiments and Detectors for a Relativistic Heavy Ion 
Collider, Brookhaven National Laboratory, July 2-7, 1990,
Editors: M. Fatyga and B. Moskowitz, pp. 169-206 (BNL, 1990).

[6]  L. Van Hove. Z. Phys. C. Particles and Fields 21,93-98 (1983),
Hadronization Model Quark-Gluon Plasma in Ultra-Relativistic Collisions 
CERN-TH 3924 (1984).

[7]  L. Van Hove  Nucl. Phys. A46 (1987).

[8]  M. Gyulassy, H. Kajantie, Kurki-Suuno and L. McLerran, Nucl.
Phys. B237 (1984) 477.

[9]  F.E. Paige and S.D. Protopopescu, "ISAJET" A MONTE CARLO event
generator program for pp and p-bar + p interactions; BNL-29777 (1991);
BNL-31987, Sept. 1982 and modifications.

[10] R.D. Field and R.P. Feynman, Nuci. Phys. B136, 1 (1978).

[11]  T. Ludlam, et al., RHIC Workshop I, eds. P. Haustein and C.L. Woody
(Brookhaven, April 85).

[12] X.N. Wang and M. Gyulassy.  Phys. Rev. D 44 (1991) 3501; Phys. Rev.
D 45 (1992) 844; Comp. Phys. Comm. 83 (1994) 307.

[13] B. Anderson, et al., Nucl. Phys. B 281, 289 (1987); Comp. Phys.
Comm. 43, 387 (1987).

[14] T. Sjostrand, Comp. Phys. Comm 82, 74 (1994).

[15] M.Gyulassy, Nuclear Physics A 590 (19995) 431c-446c.

[16] A. Capella, U. Sukhatme, C. I. Tan and J. Tran Thanh Van, Phys. Rep.
236, 225 (1994).

[17] K. Werner, Phys. Rev. 282, 87 (1993).

[18] H. Sorge, H. Stocker, and W. Greiner, Nucl. Phys. A 498, 567c (1989).

[19] H. Sorge, H. Stocker and W. Greiner, Nucl. Phys. A 566, 663c (1994).

[20] K. Geiger and B Mueller, Nucl. Phys. B 369, 600 (1992).

[21] P. Koch, B. M\"{u}ller and J. Rafelski, Phys. Rep. 142, 167 (1986).

[22] L. D. Landau, Izu. Akad. Nauk SSSR 17, 51 (1953).

[23] E. Schnedermann and U. Heinz, Phys. Rev. Lett. 69, 2908 (1992).

[24] JACEE Collaboration, Phys. Rev. Lett. 50, 2062 (1983).

[25] C.M.G. Lattes, Y. Fujimoto and S. Hasegawa, Phys. Rep. 65, 151
(1980); L.T. Baradzei et al., Nucl. Phys. B370, 365 (1992).

[26] K. Rajagopal and F. Wilczek, Nucl. Phys. B379, 395 (1993).

[27] We have chosen the 3 direction in isospin to be the neutral pions.

[28] T.J. Hallman, et al. (STAR Collaboration).  The Physics and
Detectors of the Relativistic Heavy Ion Collider (RHIC).  Proc. XXXIst 
Rencontres de Moriond, Les Arcs. Savoie, France, March 23-30, 1996.  
Series:  Moriond Particle Physics Meetings, '96 QCD and High Energy 
Hadronic Interactions, edited by J. Tran Thanh Van, published by 
Editions Frontieres, pp. 485-90 (1996).

[29] Klaus Geiger, Ron Longacre, and Dinesh K. Srivastava, VNI;Simulation
of High-Energy Particle Collisions in QCD, BNL-65755 (1998), Klaus Geiger
and Ron Longacre Heavy Ion Physics 8, 41(1998).

[30] D. Kharzeev and R. Pisarski, hep-ph/9906401, January 16, 1999.

[31] D. Kharzeev, private communication.

[32] M. Gyulassy, RBRC Memo, 3/11/99.

[33] R. Longacre,  Parton structure through two particle correlations in 
Au-Au at RHIC.  Proc. RHIC Physics and Beyond:  Kay Kay Gee Day, October 23,
1998, Brookhaven National Laboratory, Upton, NY (in press).

\end {document}